\documentclass[aps,prd,10pt,twocolumn,superscriptaddress,nofootinbib]{revtex4-2}
\usepackage{titletoc}
\titlecontents{section}
[1cm] 
{\addvspace{1pt}} 
{\contentslabel{2em}} 
{} 
{\titlerule*[0.5pc]{.}\contentspage} 

\titlecontents{subsection}
[1.5cm] 
{\addvspace{0pt}} 
{\contentslabel{1.5em}} 
{} 
{\titlerule*[0.5pc]{.}\contentspage} 

\usepackage{mathrsfs, amssymb, amsmath, amsfonts, 
	latexsym, graphicx}  
\usepackage[utf8]{inputenc}

\usepackage[hang,flushmargin]{footmisc} 
\usepackage[
colorlinks=true,
linkcolor=blue,
urlcolor=cyan,
filecolor=magenta,
citecolor=purple,
pdfstartview=FitV,
pdftitle={Carroll Schrodinger Equation},
pdfauthor={Mojtaba Najafizadeh},
pdfsubject={Carroll Schrodinger Equation},
pdfkeywords={Carroll Schrodinger Equation},
bookmarksopen=true
hyperfootnotes=None
]{hyperref}

\usepackage[all]{hypcap}

\usepackage{soul}

\usepackage{lipsum}

\usepackage{tikz}
\usepackage{xcolor}

\definecolor{galilei}{RGB}{255,255,153}
\definecolor{minkowski}{RGB}{204,255,204}
\definecolor{carroll}{RGB}{255,204,204}

\usetikzlibrary{calc}

\newcommand\boxedB[1]{{\setlength\fboxsep{8pt}\boxed{#1}}}

\usepackage{upgreek}

\usepackage{footnotebackref}

\usepackage{mathtools}

\DeclareFontFamily{U}{mathx}{\hyphenchar\font45 }
\DeclareFontShape{U}{mathx}{m}{n}{
	<-> mathx10
}{}
\DeclareSymbolFont{mathx}{U}{mathx}{m}{n}
\DeclareMathAccent{\Widetilde}{0}{mathx}{"72}

\def\be {\begin{equation}}
	\def\ee {\end{equation}}
\def\le {\left}
\def\ri {\right}
\def\p {\partial}
\def\d {\delta}
\def\m {\mu}

\begin{document}

	
	
	\title{{\LARGE Carroll-Schr\"odinger Equation}\vspace{2pt}\\\noindent\rule{250pt}{1.3pt}\vspace{5pt}}
	
	\author{Mojtaba \textsc{Najafizadeh}\,\vspace{1pt}}

	
	\affiliation{School of Physics, Institute for Research in Fundamental Sciences (IPM), \\P.O.Box 19395-5531, Tehran, Iran \\[10pt]}

	\affiliation{Department of Physics, Faculty of Science, Ferdowsi University of Mashhad, \\ P.O.Box 1436, Mashhad, Iran \\
		\vspace{.3cm} \href{mailto:mnajafizadeh@ipm.ir}{\color{blue} mnajafizadeh@ipm.ir} \vspace{.2cm}}

	\begin{abstract}
		\noindent{\bf Abstract:} The Poincar\'e symmetry can be contracted in two ways to yield the Galilei symmetry and the Carroll symmetry. The well-known Schr\"odinger equation exhibits the Galilei symmetry and is a fundamental equation in Galilean quantum mechanics. However, the question remains: what is the quantum equation that corresponds to the Carroll symmetry? In this paper, we derive a novel equation in two dimensions, called the ``Carroll-Schr\"odinger equation'', which describes the quantum dynamics in the Carrollian framework. We also construct the so-called ``Carroll-Schr\"odinger algebra'' in two dimensions, which is a conformal extension of the centrally extended Carroll algebra with a dynamical exponent of $z=1/2$. We demonstrate that this algebra is the symmetry algebra of the Carroll-Schr\"odinger field theory. Moreover, we apply the method of canonical quantization to the theory and utilize it to compute the transition amplitude. Finally, we discuss higher dimensions and identify the so-called ``generalized Carroll-Schr\"odinger equation''.  
		
		\vspace{.2cm}
	\end{abstract}

	\maketitle

	\onecolumngrid
	{\small \tableofcontents}
	
	\newpage
	
	\twocolumngrid


	\section{Introduction}
	
	In light of the inadequacy of the Galilei symmetry to describe electrodynamics, it was necessary to extend it to the Lorentz symmetry, or more generally, the Poincar\'e symmetry. Through a process of contraction, it was anticipated that the Poincar\'e symmetry would reproduce the Galilei symmetry. This contraction is characterized by small characteristic velocities compared to the speed of light, referred to as the $c\to \infty$ limit or the Galilei/nonrelativistic limit.

	Alternatively, the Poincar\'e symmetry can be contracted by considering characteristic velocities that are much larger than the speed of light, known as the $c\to 0$ limit or the Carroll/ultrarelativistic limit. This distinct type of contraction was discovered by L\'evy-Leblond in 1965 and is known as the Carroll symmetry \cite{Levy1965}.

	Both the Galilei and the Carroll symmetries are two distinct contractions of the Poincar\'e symmetry, depending on the limit of the speed of light. In terms of the spacetime coordinates, the limit can be expressed as $x_i\to \epsilon \,x_i$, $t\to t$, $\epsilon\to 0$ for the Galilei case, and $x_i\to x_i$, $t\to \epsilon \,t$, $\epsilon\to 0$ for the Carroll case.

	While these two contractions are taken from the same thing, it is worth noting that the Galilei symmetry aligns with empirical evidence observed in the natural world, whereas the Carroll symmetry represents a theoretical framework that has the potential to uncover novel aspects of physics.

	The study of Carrollian structures and their application to diverse physical systems is driven by several motivations. Notably, a recent motivation that has stimulated this field of research is the fact that the Carrollian conformal algebra is isomorphic to the BMS algebra in one dimension higher \cite{Duval:2014uoa, Duval:2014uva, Duval:2014lpa} and therefore is relevant for the celestial approach to flat space holography  \cite{Barnich:2009se,Bagchi:2010zz,Ciambelli:2018wre,Figueroa-OFarrill:2021sxz,Herfray:2021qmp,Bagchi:2022emh,Donnay:2022aba,Campoleoni:2022wmf,Donnay:2022wvx,Bagchi:2023fbj,Saha:2023hsl,Salzer:2023jqv,Campoleoni:2023fug,Nguyen:2023vfz}.

	Other aspects of Carrollian physics has been studied by different motivations, see e.g. 
	\cite{Henneaux:1979vn,Dautcourt:1997hb,
		Bergshoeff:2014jla,Nzotungicimpaye:2014wya,
		Bergshoeff:2015wma,Bekaert:2015xua,Hartong:2015xda,
		Bergshoeff:2017btm,Duval:2017els,deBoer:2017ing,
		Ciambelli:2018xat,Ciambelli:2018ojf,Morand:2018tke,Penna:2018gfx,
		Donnay:2019jiz,Bergshoeff:2019ctr,Ravera:2019ize,Gomis:2019nih,Ciambelli:2019lap,Ballesteros:2019mxi,
		Bergshoeff:2020xhv,Niedermaier:2020jdy,Gomis:2020wxp,Grumiller:2020elf,
		Hansen:2021fxi,deBoer:2021jej,Henneaux:2021yzg,Concha:2021jnn,Guerrieri:2021cdz,Perez:2021abf,
		Figueroa-OFarrill:2022mcy,Campoleoni:2022ebj,Baiguera:2022lsw,Perez:2022jpr,Fuentealba:2022gdx,Marsot:2022imf,Bergshoeff:2022qkx,Banerjee:2022ocj,Bagchi:2022eui,Bagchi:2022owq,Bekaert:2022oeh,Rivera-Betancour:2022lkc,deBoer:2023fnj,Ecker:2023uwm,Koutrolikos:2023evq,Kasikci:2023zdn,Ciambelli:2023tzb,Ciambelli:2023xqk,Bergshoeff:2023vfd} and references therein. In particular, it could potentially have implications for de Sitter cosmology and the phenomenon of inflation \cite{deBoer:2021jej,deBoer:2023fnj}, as well as the Hall effect \cite{Marsot:2022imf}, hydrodynamics \cite{deBoer:2017ing,Ciambelli:2018xat,Ciambelli:2018wre}, scalar fields \cite{Henneaux:2021yzg,deBoer:2021jej,Bergshoeff:2022qkx,Rivera-Betancour:2022lkc}, fermionic fields \cite{Banerjee:2022ocj,Bagchi:2022eui,Koutrolikos:2023evq,Bergshoeff:2023vfd}, and supersymmetry \cite{Bergshoeff:2015wma,Bagchi:2022owq,Koutrolikos:2023evq}.

	The Carrollian physics could offer intriguing possibilities, particularly in the context of quantum mechanics. Usually, the Schr\"odinger equation can be obtained from the Klein-Gordon equation by applying a field redefinition and taking the Galilei/nonrelativistic limit $c\to\infty$ (see e.g. \cite{Leblanc:1992wu} or a brief review in Appendix \ref{ap}). However, in the Carrollian sector, we find a similar approach proves unsuccessful when starting from the Klein-Gordon equation. Instead, one has to begin with the tachyon Klein-Gordon equation. 
	
	Consequently, the main objective of this paper is to identify a Carrollian quantum equation, which we call the ``Carroll-Schr\"odinger equation'', analogous to the Schr\"odinger equation. This is investigated in Section \ref{II} for the case of two dimensions.
	
	

	The Schr\"odinger algebra is a conformal extension of the Galilei algebra with a central charge. The field theories with the Schr\"odinger symmetries have been studied in \cite{Jackiw:1990mb,Henkel:1993sg,Nishida:2007pj}. This algebra captures the symmetry of the Schr\"odinger equation.

	Similarly, another goal of this work is to construct the so-called ``Carroll-Schr\"odinger algebra'' in two dimensions, which is a conformal extension of the Carroll algebra with a central charge. Moreover, we illustrate the existence of an infinite-dimensional extension of this algebra. These developments are detailed in Section \ref{III}.

	Furthermore, as detailed in Section \ref{inv}, we demonstrate that our two-dimensional Carrollian quantum equation exhibits the Carroll-Schr\"odinger algebra as its underlying symmetry.

	Additionally, we delve deeper into the Carroll-Schr\"odinger field theory in Section \ref{V} by applying the method of canonical quantization to the theory and computing the transition amplitude. 
	
	Finally, in Section \ref{2}, we discuss higher dimensions and derive an equation, called the ``generalized Carroll-Schr\"odinger equation'', which, in two dimensions, reduces to the Carroll-Sch\"odinger equation. 
	
	Therefore, throughout this work, while the ``generalized Carroll-Sch\"odinger'' refers to an equation/action in higher dimensions, the ``Carroll-Sch\"odinger'' specifically denotes an equation/action/algebra in two dimensions.

	The shorthand notations $\p_t:=\p/\p t$, $\p_x:=\p/\p x$, and $\p_i:=\p/\p x^i$ (with $i=1,\ldots,d$) are used.

	\section{Carroll-Schr\"odinger equation} \label{II}
	Let us take into account the Klein–Gordon equation for a complex tachyon field $\phi$ of mass $m$ in a two-dimensional Minkowski spacetime, with the mostly plus signature for the metric, including the speed of light $c$ and the reduced Planck constant $\hbar$
	\begin{align}
		\le(-\,\frac{1}{c^2}\,\p_t^2\,+\,\p_x^2\,+\,\m^2 \ri)\phi=0\,,\label{KG}
	\end{align} 
	where $\m:={mc}/{\hbar}$. By applying a field redefinition
	\begin{align}
		\phi=\frac{1}{\sqrt{\m}}~e^{-\,i\m x}\,\psi\,, \label{fieldred}
	\end{align}
	the tachyon Klein-Gordon equation \eqref{KG} reduces to
	\begin{align}
		\le(-\,\frac{1}{c^2}\,\p_t^2\,-\,2i\m\,\p_x\,+\,\p_x^2\ri)\psi=0\,.\label{reduced}
	\end{align}
	Using a dimensionless parameter $\epsilon$, we can first rescale $\m\to\m/\epsilon^2$ in \eqref{reduced} and then take the Carroll limit
	\be 
	x\to x\,, \qquad t\to\epsilon\,t\,,\qquad \epsilon\to 0\,.\label{carrlimi}
	\ee
	Through this process, we obtain the following equation  
	\begin{align}
		\boxedB{\le(i\hbar\,c\,\p_x+\frac{\hbar^2}{2m c^2}\,\p_t^2 \ri)\psi=0\,.}\label{cse}
	\end{align}
	We will refer to this equation as the ``Carroll-Schr\"odinger equation'', because it has a Carroll-Schr\"odinger symmetry, as will be discussed later.

	The operator acting on the field $\psi$ in \eqref{cse} is Hermitian with respect to the Hermitian conjugation rules: $(\p_t)^\dagger\equiv-\,\p_t$, and $(\p_x)^\dagger\equiv-\,\p_x$. This suggests that the Carroll-Schr\"odinger equation \eqref{cse} can be derived from an action principal, which we call the ``Carroll-Schr\"odinger action''
	\begin{align}
		\boxedB{S=\int dt\,dx\,\psi^\star\le(i\hbar\,c\,\p_x+\frac{\hbar^2}{2m c^2}\,\p_t^2 \ri)\psi\,.}\label{csa}
	\end{align}
	We note that the introduced $\mu$ in \eqref{KG} has the length dimension $[\mu]_{_L}=-1$. In two spacetime dimensions, the length dimension of a scalar field is $[\phi]_{_L}=0$. Hence, the coefficient in the field redefinition \eqref{fieldred} is chosen to ensure that the length dimension of the Carroll-Schr\"odinger field $\psi$ becomes $[\psi]_{_L}=-1/2$. This choice ensures that the action \eqref{csa} possesses the expected dimension of $\hbar$.

	It is worth noting that when the above procedure is followed and a real scalar field $\phi$ is employed, such as $\phi\propto e^{-\mu x}\,\psi$, one arrives at an equation that the operator acting on the field $\psi$ is not Hermitian, and thus cannot be derived from an action.

	Moreover, if we begin with a complex scalar field equation with a real mass, rather than a complex tachyon field equation \eqref{KG}, it would result in an equation that becomes ill-defined in the Carroll limit \eqref{carrlimi}. This highlights the necessity of initiating the analysis with a tachyon field equation \eqref{KG}.

	We note that it is possible to begin with a field redefinition \eqref{fieldred} by considering the opposite sign of $\mu$. This would lead to the replacement of $m\to -\,m$ in the resulting equation and action, presenting another possibility.

	In our approach, we first derived the equation of motion and subsequently obtained the action. We notice that by beginning with the action of a complex tachyon field, applying the field redefinition \eqref{fieldred}, and taking the Carroll limit, we could directly arrive at the action \eqref{csa}.

	It is also important to emphasize that taking the limit is a technique to obtain a new physics. Accordingly, after taking the limit \eqref{carrlimi}, the nature of initial field has been changed. Instead of being a tachyon, it now represents a massive Carrollian field.

	\section{Carroll-Schr\"odinger algebra} \label{III}
	In this section, we introduce the ``Carroll-Schr\"odinger algebra'' in two spacetime dimensions, including both finite-dimensional and infinite-dimensional extensions.

	\subsection{Finite-dimensional extension}
	The Carroll algebra in $1+1$ dimensions, involving time translation $H$, space translation $P$, and Carroll boost $B$, is given by $[\,P\,,\,B\,]=H$. It is convenient to see that this algebra admits two non-trivial central extensions as $[\,H\,,\,B\,]=c_1$, and $[\,H\,,\,P\,]=c_2$. By ignoring $c_2$, and identifying $c_1$ with mass parameter, $c_1:=M$, the double centrally extended Carroll algebra reduces to
	\be 
	\boxedB{~~~
		\begin{aligned}
			&[\,P\,,\,B\,]=H\,,   \qquad  && [\,H\,,\,B\,]=M\,. \label{carrext}
		\end{aligned}
		~~~}
	\ee 
	We name this the ``Carroll-Bargmann algebra'', denoted by $\mathfrak{carrb}(1+1)$, drawing an analogy to the Bargmann algebra which is a central extension of the Galilei algebra.

	Next, we extend the Carroll-Bargmann algebra \eqref{carrext} by including the conformal generators, namely the dilatation $D$ and the spatial special conformal transformation $K$. We find such an extended algebra, that is a $2$-dimensional conformal Carroll algebra with the central charge $M$, has the following nonzero brackets
	\be 
	\boxedB{~~~
		\begin{aligned}
			&[\,P\,,\,B\,]=H\,,       \quad & &[\,H\,,\,B\,]={M}\,,\\[2pt]
			&[\,P\,,\,D\,]=2\,P\,, \quad & &[\,D\,,\,K\,]=2\,K\,, \\[2pt]
			&[\,H\,,\,D\,]=H\,, \quad & &[\,P\,,\,K\,]=D\,, \\[2pt]
			&[\,D\,,\,B\,]=B\,, \quad & &[\,H\,,\,K\,]=B\,.\label{csch}
		\end{aligned}
		~~~}
	\ee
	We will refer to this algebra as the ``Carroll-Schr\"odinger algebra'', denoted by $\mathfrak{carrsch}(1+1)$, because it is the symmetry group of the Carroll-Schr\"odinger action \eqref{csa}, as will be shown in the next section.

	The generators of the Carroll-Schr\"odinger algebra \eqref{csch} can be represented by
	\begin{align}
		&H=\p_t\,,\qquad &  &B=x\,\p_t-i m\,t\,,  \label{generators} \\[5pt]
		&P=\p_x\,,\qquad & &D=t\,\p_t+2\,x\,\p_x+\omega\,,  \nonumber\\[5pt]
		&{M}=-\,i m\,, \qquad & &K=x\,t\,\p_t+x^2\,\p_x-\tfrac{1}{2}\,i m \,t^2+\omega x\,,\nonumber
	\end{align}
	where $\omega$ is the dilatation weight.

	We note that, as demonstrated in \eqref{generators}, the dilatation generator $D$ scales space and time differently $x\to\lambda^2 x$, $t\to\lambda \,t$, indicating a critical exponent of $z=1/2$. However, the corresponding generator in the Carrollian conformal algebra \cite{Duval:2014lpa} scales space and time in the same way $x \to \lambda \,x$, $t \to \lambda \,t$, with a critical exponent of $z = 1$.
	
	\subsection{Infinite-dimensional extension}
	
	The $z=1/2$ Carroll-Schr\"odinger algebra \eqref{csch} can be made isomorphic to the $z=2$ Schr\"odinger algebra by replacing the generators $P$ and $H$ with each other. This allows us to simply write an infinite-dimensional extension of the algebra \eqref{csch}.

	
	Accordingly, we can present the infinite-dimensional Carroll-Schr\"odinger algebra, denoted as $\mathfrak{\Widetilde{carrsch}}(1+1)$, as
	\be 
	\boxedB{~~~
		\begin{aligned}
			&[\,L_n\,,\,L_m\,] = (n-m)\,L_{n+m}\,, \\[7pt]
			&[\,L_n\,,\,Y_\ell\,] = \le(\,\tfrac{n}{2}-\ell\,\ri)Y_{n+\ell}\,, \\[7pt]
			&[\,M_n\,,\,L_m\,]=n\,M_{n+m}\,,\\[7pt]
			&[\,Y_\ell\,,\,Y_{k}\,] = (\ell -k)\,M_{\ell+k}\,,\\[7pt]
			&[\,Y_\ell\,,\,M_{n}\,]=0=[\,M_n\,,\,M_m\,]\,, \label{infcs}
		\end{aligned}
		~~~}
	\ee
	where $ n,m\in\mathbb{Z}$, and $\ell,k\in\mathbb{Z}+\tfrac{1}{2}$. The infinite-dimensional generators that fulfill the algebra \eqref{infcs} can be given by	
	\begin{align}
		\!\!	L_n&=-\,x^{n+1}\,\p_x-\,\tfrac{1}{2}\,(n+1)\,x^n\,t\,\p_t \nonumber\\[7pt]
		\!\!	&~~~  \,-\tfrac{1}{4}\,M\,n(n+1)\,t^2\,x^{n-1}-\,\tfrac{1}{2}\,\omega\,(n+1)\,x^n\,, \\[7pt] 
		\!\!    Y_\ell&=-\,x^{\,(\ell+\frac{1}{2})}\,\p_t-\,M\,(\ell+\tfrac{1}{2})\,t\,x^{(\ell-\frac{1}{2})}\,,\label{yell}\\[7pt]
		\!\!	M_n&=-\,M\,x^n\,,
	\end{align}
	where $\omega$ is the dilatation weight, and $M$ is an arbitrary constant representing the mass parameter.

	It is understood that the finite-dimensional generators are those with $M_0$, $Y_{\pm \frac{1}{2}}$, $L_{\pm 1,0}$. In other words, we can write the generators in \eqref{generators} in a suggestive form	
	\begin{align}
		&M=-\,im=-\,M_0\,, \nonumber   \\[5pt]
		&B=x\,\p_t-im\,t=-\,Y_{\frac{1}{2}}\,,  \nonumber\\[5pt]
		&H=\p_t=-\,Y_{-\frac{1}{2}}\,, \nonumber   \\[5pt]
		&P=\p_x=-\,L_{_{-1}}\,,	       \nonumber\\[5pt]
		&D=t\,\p_t+2\,x\,\p_x+\omega=-\,2L_{_0}\,, \nonumber\\[5pt]	
		&K=x\,t\,\p_t+x^2\,\p_x-\tfrac{1}{2}\,i m \,t^2+\omega x=-\,L_{_1}\,.	 \label{generatorsi}
	\end{align}
	As a result, the finite-dimensional Carroll-Schr\"odinger algebra \eqref{csch} would be a sub-algebra of \eqref{infcs}, leading to the following hierarchy of Lie subalgebras:
	\be 
	\mathfrak{carr}(1+1)\subset\mathfrak{carrb}(1+1)\subset\mathfrak{carrsch}(1+1) \subset\mathfrak{\Widetilde{carrsch}}(1+1)\,. \nonumber
	\ee
	
	We note that, in two spacetime dimensions, the $z=2$ Schr\"odinger algebra and the $z=\frac{1}{2}$ Carroll-Schr\"odinger algebra indeed share the same infinite-dimensional structure, analogous to that of the Carrollian conformal algebra (CCA) and the Galilean conformal algebra (GCA). We also note that the triplet $P$, $D$, $K$ (or $L_{-1}$, $L_0$, $L_1$), forms an $sl(2;R)$ subalgebra.

	\section{Symmetries of the action} \label{inv}
	
	This section demonstrates that the Carroll-Schr\"odinger algebra \eqref{csch} is the symmetry algebra of the Carroll-Schr\"odinger action \eqref{csa}, while the infinite-dimensional Carroll-Schr\"odinger algebra \eqref{infcs} does not. This is what we demonstrate in this section, so readers may proceed to the next section without any loss of generality.

	The transformation of the Carroll-Schr\"odinger field can be given by
	\begin{align}
		\d\psi&= \big(~\lambda_{_H}\,H~+~\lambda_{_P}\,P~+~\lambda_{_B}\,B\nonumber\\[3pt]&\,\quad~\quad+~\lambda_{_M}\,M~+~\lambda_{_D}\,D~+~\lambda_{_K}\,K~\big)\,\psi\,,\label{transf}
	\end{align} 
	where $H$, $P$, $B$, $M$, $D$, $K$ are the generators of the Carroll-Schr\"odinger algebra \eqref{csch}, represented in \eqref{generators}, and $\lambda_{_H}$, $\lambda_{_P}$, $\lambda_{_B}$, $\lambda_{_M}$, $\lambda_{_D}$, $\lambda_{_K}$ are the corresponding parameters associated with each transformation.

	Accordingly, we can demonstrate that the Carroll-Schr\"odinger action is invariant under the transformation \eqref{transf}, upon the dilatation weight $\omega=1/2$. This invariance implies that the Carroll-Schr\"odinger action \eqref{csa}, and consequently the equation \eqref{cse}, have the Carroll-Schr\"odinger algebra \eqref{csch} as their symmetry algebra.

	To illustrate this, assuming $c=1=\hbar$, let us consider the action \eqref{csa} as
	\be 
	S=\int dt\,dx~\psi^\dagger\,\mathbb{K}\,\psi\,, \label{ac2}
	\ee  
	where $\mathbb{K}$ is defined as the ``Carroll-Schr\"odinger operator''
	\be 
	\mathbb{K}:=i\,\p_x+\frac{1}{2m}\,\p_t^2\,.
	\ee 
	The variation of the action \eqref{ac2} gives
	\be 
	\d S = \int dt\,dx\,\le(\d\psi^\dagger\,\mathbb{K}\,\psi+\psi^\dagger\,\mathbb{K}\,\d\psi\ri)\,,
	\ee 
	which requires knowing the Hermitian conjugate of $\d\,\psi$; i.e. $\d\,\psi^\dagger$. For this purpose, we first find the Hermitian conjugates of the generators in \eqref{generators}. These can be found by applying the Hermitian conjugation rules
	\be 
	(\p_t)^\dagger\equiv-\,\p_t\,,\quad\quad (\p_x)^\dagger\equiv-\,\p_x\,,\quad\quad t^\dagger\equiv t\,,\quad\quad x^\dagger\equiv x
	\ee  
	to the Carroll-Schr\"odinger generators \eqref{generators}, resulting in
	\begin{align}
		&H^\dagger=-\,\p_t\,, &  &B^\dagger=-\,x\,\p_t+i m\,t\,,  \label{generators--} \\[8pt]
		&P^\dagger=-\,\p_x\,, & &D^\dagger=-\,t\,\p_t-2\,x\,\p_x+\omega-3\,,  \nonumber\\[8pt]
		&M^\dagger=im\,,  & &K^\dagger=-\,x\,t\,\p_t-x^2\,\p_x+\tfrac{1}{2}\,i m \,t^2+ x\,(\omega-3)\,.\nonumber
	\end{align}
	As a result, using the latter, we can express the Hermitian conjugation of $\d\psi$ as follows
	\begin{align}
		\d\psi^\dagger&=\psi^\dagger\, \big(~\lambda_{_H}\,H^\dagger~+~\lambda_{_P}\,P^\dagger~+~\lambda_{_B}\,B^\dagger\nonumber\\[3pt]&\,\quad~\quad~~~~+~\lambda_{_M}\,M^\dagger~+~\lambda_{_D}\,D^\dagger~+~\lambda_{_K}\,K^\dagger~\big)\,.
	\end{align} 
	Now, we can demonstrate the invariance of the action \eqref{ac2} under each transformation. For example, under the dilatation $D$, we have
	\begin{align}
		\d_{_D} S&=\int dt\,dx\,\le(\d_{_D}\psi^\dagger\,\mathbb{K}\,\psi+\psi^\dagger\,\mathbb{K}\,\d_{_D}\psi\ri)\nonumber\\[5pt]
		&=\int dt\,dx\,\lambda_{_D}\psi^\dagger\Big[\, D^\dagger\, \mathbb{K}+ \mathbb{K}\, D\,\Big]\psi\nonumber\\[5pt]
		&=\int dt\,dx\,\lambda_{_D}\psi^\dagger\big(2\,\omega-1\big)\,\mathbb{K}\,\psi\,,
	\end{align}
	which vanishes for $\omega=1/2$. Moreover, under the spatial special conformal transformation $K$, we find
	\begin{align}
		\d_{_K} S&=\int dt\,dx\,\le(\d_{_K}\psi^\dagger\,\mathbb{K}\,\psi+\psi^\dagger\,\mathbb{K}\,\d_{_K}\psi\ri)\\[5pt]
		&=\int dt\,dx\,\lambda_{_K}\psi^\dagger\!\le[
		x\,(2\,\omega-1\big)\,\mathbb{K}+i\le(\omega-\tfrac{1}{2}\ri)\ri]\psi\,,\nonumber
	\end{align}
	which again vanishes for $\omega=1/2$. Similarly, it is straightforward to demonstrate the invariance of the action \eqref{ac2} under other transformations.

	Let us now examine the invariance of the action under the transformations associated with the generators of the infinite-dimensional algebra \eqref{infcs}. For instance, consider the transformation of the field under the generator $Y_{\ell}$, given by \eqref{yell},  
	\be 
	\delta_{_Y}\psi=\lambda_{_Y}\,Y_{\ell}~\psi\,,\label{dely}
	\ee 
	where $\lambda_{_Y}$ is the transformation parameter. For simplicity, let us examine the invariance at the level of the equation. By varying the Carroll-Schr\"odinger equation \eqref{cse}, assuming $c=1=\hbar$, under the transformation \eqref{dely}, $(i\,\p_x+\tfrac{1}{2m}\,\p_t^2)\,\delta_{_Y}\psi=0$, we obtain
	\be 
	\lambda_{_Y}\!\le[-\,m(\ell-\tfrac{1}{2})(\ell+\tfrac{1}{2})\,t\,x^{(\ell-\frac{3}{2})}\ri]\psi\approx0\,,\label{weak}
	\ee 
	where the ``weak equality'' symbol $\approx$ indicates that the equation of motion is applied to reach \eqref{weak}. This relation demonstrates that the differential equation is not invariant under the transformation \eqref{dely}. However, it could be invariant if we choose $\ell=\pm \,\frac{1}{2}$, corresponding to the invariance under the finite-dimensional generators, i.e. the boost $B=-Y_{\frac{1}{2}}$ and the time translation $H=-Y_{-\frac{1}{2}}$. Similarly, we can illustrate that the differential equation would not be invariant under $L_n$ and $M_n$, except for the finite-dimensional generators $L_{\pm 1,0}$ and $M_0$.

	As a result, it is found that only the finite-dimensional Carroll-Schr\"odinger algebra \eqref{csch} is the symmetry algebra of the Carroll-Schr\"odinger equation/action \eqref{cse}/\eqref{csa}, while the infinite-dimensional algebra \eqref{infcs} does not.

	\section{Further study on Carroll-Schr\"odinger} \label{V}

	It is interesting to study the Carroll-Schr\"odinger field theory in more detail. In this section, we apply the method of canonical quantization and subsequently use it to compute the transition amplitude.

	\subsection{Canonical quantization}\label{fq}

	The field quantization requires us to consider it as a quantum system rather than a classical one. Therefore, in two spacetime dimensions, we consider the Lagrangian density of the Carroll-Schr\"odinger field theory as
	\be 
	\mathcal{L}=\tfrac{1}{2}\,i\hbar c\,\Big[\psi^\dagger\p_x\psi-(\p_x\psi^\dagger)\psi\Big]+\frac{\hbar^2}{2mc^2}\,\psi^\dagger\p_t^2\,\psi\,,\label{ld}
	\ee 
	where the fields $\psi$ and $\psi^\dagger$ are treated as operators. We note that this form of the Lagrangian density is indeed equivalent to the one in \eqref{csa} up to a total derivative. The advantage of this form is that we can now introduce canonical momenta conjugate to both fields $\psi$ and $\psi^\dagger$, as derived below. In contrast, using the previous form in \eqref{csa}, only one canonical momentum can be defined.

	A general solution of the Carroll-Schr\"odinger equation for $\psi$, and its complex conjugate for $\psi^\dagger$, can be given by the plane-wave solutions
	\begin{align} 
		\psi(x,t)&=\int\,\frac{dE}{2\pi\hbar c}~a(E)\,e^{-\,i\le(Px-Et\ri)/\hbar}\,,\label{ps1}\\[5pt]
		\psi^\dagger(x,t)&=\int\,\frac{dE}{2\pi\hbar c}~a^\dagger(E)\,e^{i\le(Px-Et\ri)/\hbar}\,,\label{ps2}
	\end{align}
	where the coefficients $a(E)$ and $a^\dagger(E)$ are also operators with the length dimension of $1/2$, and
	\be 
	P=\frac{E^2}{2mc^3}\label{em}
	\ee
	is the energy-momentum relation in the Carrollian framework, as detailed in \cite{Najafizadeh:2024}. We note that negative energies are possible within the Carrollian framework, since it was derived from a tachyonic scenario. This can also be confirmed by \eqref{em}, where momentum is positive definite, but energy can be negative. This is in contrast to the Galilean case, where energy is positive definite, $E=P^2/2m$, while momentum can be negative.

	We then define new canonical momenta conjugate to the fields $\psi$ and $\psi^\dagger$, subject to their spatial derivatives, as follows (a similar definition was used in \cite{Koutrolikos:2023evq}):
	\begin{align} 
		\pi&:=\frac{\p\mathcal{L}}{\p\mathring{\psi}}=i\hbar c\,\psi^\dagger\,, ~~~~~~\qquad \mathring{\psi}\equiv \tfrac{1}{2}\, \p_x\psi\,,\label{pi1}\\[3pt]
		\pi^\dagger&:=\frac{\p\mathcal{L}}{\p\mathring{\psi^\dagger}}=-\,i\hbar c\,\psi\,, ~~\qquad \mathring{\psi^\dagger}\equiv \tfrac{1}{2}\, \p_x\psi^\dagger\,.\label{pi2}
	\end{align}
	Accordingly, in the Carrollian framework, we can present the so-called ``equal-position commutation relations''
	\begin{align} 
		[\,\psi(x,t)\,,\,\pi(x,t')\,]&=i\hbar\,\d(t-t') \label{commcar}\\[3pt]
		[\,\psi(x,t)\,,\,\psi(x,t')\,]&=[\,\pi(x,t)\,,\,\pi(x,t')\,]=0\,, \nonumber
	\end{align} 
	and similar relations for their Hermitian conjugates. By applying \eqref{ps1} and \eqref{pi1} in the quantization conditions \eqref{commcar}, we can conveniently find the commutation relations between the operators $a$ and $a^\dagger$, which become
	\begin{align} 
		&[\,a(E)\,,\,a^\dagger(E')\,]=2\pi\hbar\,c\,\d(E-E')\,,\label{commcar2}\\[3pt]
		&[\,a(E)\,,\,a(E')\,]=[\,a^\dagger(E)\,,\,a^\dagger(E')\,]=0\,. \nonumber
	\end{align}
	These commutation relations can also be satisfied by
	\begin{align} 
		a(E)&=\int c \,dt~\psi(x,t)~e^{i\le(Px-Et\ri)/\hbar}\,,\label{a1}\\[5pt]
		a^\dagger(E)&=\int c\, dt~\psi^\dagger(x,t)~e^{-\,i\le(Px-Et\ri)/\hbar}\,.\label{a2}
	\end{align}
	When we posit the existence of a vacuum state $|\,0\,\rangle$, such that $a(E)|\,0\,\rangle=0$, the particle picture emerges. The operator $a(E)$ annihilates a particle of energy $E$, while $a^\dagger(E)$ creates one. Therefore, $\psi$ annihilates particles and $\psi^\dagger$ creates particles.

	We can continue our discussion on canonical quantization, but let us delve into more detail in a future work. For now, with this information, we are equipped to compute the transition amplitude in the next section.
	
	\subsection{Transition amplitude}\label{tra a}

	In two spacetime dimensions, we study the transition amplitude (or two-point function) for the transition of a free particle from a spacetime position $\mathrm{X}\equiv(x^0,x)$ to a position $\mathrm{Y}\equiv(y^0,y)$, with $y^0>x^0$. The transition amplitude is given by
	\be 
	\mathbb{M}_{_{\,\mathrm{X}\to \mathrm{Y}}}=\langle \,\mathrm{Y}\,|\,\mathrm{X}\,\rangle\,,\label{ta}
	\ee 
	where $|\mathrm{X}\rangle$, $|\mathrm{Y}\rangle$ are states of a single particle at positions $X$, $Y$ correspondingly. These states are given by the action of field on the vacuum
	\be 
	|\,\mathrm{X}\,\rangle=\psi^\dagger(x^0,x)\,|\,0\,\rangle\,,\qquad 
	|\,\mathrm{Y}\,\rangle=\psi^\dagger(y^0,y)\,|\,0\,\rangle\,.
	\ee 	
	The transition amplitude \eqref{ta} is then
	\begin{align}
		\mathbb{M}_{_{\,\mathrm{X}\to \mathrm{Y}}}&=\langle\,0\,|\, \psi(y^0,y)\,\psi^\dagger(x^0,x)\, |\,0\,\rangle\nonumber\\[5pt]
		&=\int\,\frac{dE}{2\pi\hbar\,c}~e^{-\,i\le(P\Delta_x\,-\,E\Delta_t\ri)/\hbar}\,,\label{inte}
	\end{align}	
	where
	\be 
	\Delta_t=y^0-x^0\,,\qquad \Delta_x=y-x\,.
	\ee 
	To arrive this, we have employed \eqref{ps1}, \eqref{ps2}, utilized the quantization condition \eqref{commcar2}, and the fact that $a|\,0\,\rangle=0$. By substituting \eqref{em} into the integral \eqref{inte} and completing the square, it can be transformed into a Gaussian integral. Upon evaluating this Gaussian integral, the transition amplitude \eqref{inte} becomes
	\be 
	\boxedB{~
		\mathbb{M}_{_{\,\mathrm{X}\to \mathrm{Y}}}=\sqrt{\frac{mc}{2\pi i\hbar\,\Delta_x}}~\exp\le({\frac{imc^3}{2\hbar}\,\frac{(\Delta_t)^2}{\Delta_x}}\ri)\,,
	}\label{tam}
	\ee   
	which indicates a localization in space, as expected in the Carrollian framework. This result is consistent with the Schr\"odinger case, by exchanging $\Delta_x \leftrightarrow c\,\Delta_t$.

	It is worth noting that in this section, we utilized a method to quantize the Carrollian system, allowing us to derive the two-point function through field theory, as shown in \eqref{tam}. This result is consistent with the findings in Ref. \cite{Afshar:2024llh}, which employs an alternative approach that does not rely on field theory, validating our quantization method.

	\section{Comments on higher dimensions} \label{2}

	Up to this point, we have focused on two spacetime dimensions, but it becomes intriguing to explore the possibilities of extending beyond two dimensions. As discussed in previous sections, this can be approached both at the level of theory and at the level of algebra.

	{\bf At the level of theory:} We can proceed further since the tachyon Klein-Gordon equation \eqref{KG} can be considered in any dimension. Therefore, we begin with the Klein-Gordon equation applied to a complex scalar tachyon field $\phi$ in any spatial dimension
	\be 
	\le(-\,\frac{1}{c^2}\,\p_t^2\,+\,\p^i\p_i\,+\,\m^2\ri)\phi=0\,.
	\ee 
	By applying the field redefinition
	\be 
	\phi=\frac{1}{\sqrt{\m}}~e^{-\,i\m\sqrt{x^2}}\,\psi\,,\label{finany}
	\ee 
	with $x^2=x^ix_i$, we can arrive at the equation
	\be 
	\hspace{-11pt}\le(-\,\frac{1}{c^2}\,\p_t^2-2i\m\,\frac{1}{\sqrt{x^2}}\,x^j\p_j-i\m\,\frac{d-1}{\sqrt{x^2}}+\p^j\p_j\ri)\psi=0.
	\ee 
	After rescaling the parameter $\m\to\m/\epsilon^2$ in the latter and then applying the Carroll limit \eqref{carrlimi}, we reach to 
	\be 
	\boxedB{
		\begin{aligned}
			&\le(i\hbar c\,\nabla_x+\frac{\hbar^2}{2m c^2}\,\p_t^2\ri)\psi=0\,, \\[8pt]
			&\text{where}~~~~~~~\nabla_x=\frac{1}{\sqrt{x^2}}\le(x^i\p_i+\frac{d-1}{2}\ri)\,. \label{inany}
		\end{aligned}
	}
	\ee 
	When $d=1$, the operator $\nabla_x$ simplifies to the standard derivative $\nabla_x \to \partial_x$, and thus \eqref{inany} reduces to the Carroll-Schr\"odinger equation \eqref{cse}. Despite this equality in two dimensions and the fact that we have applied the Carroll limit \eqref{carrlimi} to derive the equation \eqref{inany} in higher dimensions, we refrain from naming it the ``Carroll-Schr\"odinger equation in higher dimensions''. Instead, we refer to \eqref{inany} as the ``generalized Carroll-Schr\"odinger equation''.

	We observe that the operator $\nabla_x$ has a length dimension of $[\nabla_x]_{_L} = -1$. Moreover, it is anti-Hermitian, i.e. $(\nabla_x)^{\,\dagger} = -\nabla_x$, with respect to the Hermitian conjugation already introduced. This guarantees that in \eqref{inany}, the operator acting on the field $\psi$ is Hermitian. Consequently, the equation \eqref{inany} can be derived from an action, which we call the ``generalized Carroll-Schr\"odinger action'' 
	\be 
	\boxedB{S=\int dt\,d^dx\,\psi^\star\le(i\hbar c\,\nabla_x+\frac{\hbar^2}{2m c^2}\,\p_t^2\ri)\psi\,.}
	\label{inanya}
	\ee 
	Recalling once again that in $d$ spatial dimension, the length dimension of a scalar field is $[\phi]_{_L}=(1-d)/2$, we can see from \eqref{finany} that the length dimension of the generalized Carroll-Schr\"odinger field is $[\psi]_{_L}=-\,d/2$. This satisfies the expected dimension of $\hbar$ for the action \eqref{inanya} and is the same as that of the Schr\"odinger field in any dimension (see appendix \ref{ap}).

	Once again, in our approach, we first derived the equation of motion \eqref{inany} and subsequently obtained the action. We notice that by beginning with the action of a complex tachyon field, applying the field redefinition \eqref{finany}, and taking the Carroll limit, we could directly arrive at the action \eqref{inanya}.

	The Schr\"odinger equation does not hold invariance under the transformation $t\to -\,t$. However, in the case of the equation \eqref{inany}, one finds that it is invariant under both the transformations $t\to -\,t$ and $x_i \to -\,x_i$ for $d>1$, while it is not invariant for $d=1$. This is similar to the ultrarelativistic wave equation \cite{Bergshoeff:2014jla}, $(-\,\partial_t^2-m^2)\phi=0$, which also remains invariant under these transformations in any dimension.

	We note that the Carroll-Schr\"odinger equation \eqref{cse} is mathematically identical to the Schr\"odinger equation in two spacetime dimensions when the time and spatial coordinates are exchanged ($x\leftrightarrow c\,t$). However, it is important to note that this identification only holds in two dimensions and breaks down in higher dimensions, as can be seen by comparing the generalized Carroll-Schr\"odinger equation \eqref{inany} and the $d$-dimensional Schr\"odinger equation.

	{\bf At the level of algebra:} To determine the symmetry algebra of the generalized Carroll-Schr\"odinger equation \eqref{inany}, one approach is to extend the Carroll-Schr\"odinger algebra \eqref{csch} to higher dimensions. However, we find that the Jacobi identity can only be satisfied in two dimensions. This aligns with the results in \cite{Afshar:2024llh}, which derived the Carroll-Schr\"odinger algebra only in two dimensions using a different method. Consequently, the symmetry algebra of the generalized Carroll-Schr\"odinger equation \eqref{inany} remains unknown. Extending the algebra \eqref{csch} to higher dimensions may require activating some of the previously vanished commutators, a task we plan to explore in the future.

	\section{Discussion}

	In the specific case of two spacetime dimensions, we have formulated a novel equation, which we have named the Carroll-Schr\"odinger equation \eqref{cse}. Our derivation initially stemmed from a relativistic tachyon equation; however, it is crucial to highlight that the resulting equation does not exhibit tachyonic behavior. Instead, it embodies the nature of Carrollian dynamics.

	In addition, we have successfully achieved a conformal extension of the centrally extended Carroll algebra in two dimensions. This extension, called the Carroll-Schr\"odinger algebra \eqref{csch}, could be extended to the infinite-dimensional version \eqref{infcs} demonstrating a same structure as that of the Schr\"odinger algebra. We found that the Carroll-Schr\"odinger algebra \eqref{csch} serves as the symmetry algebra for the derived action \eqref{csa}, while the infinite-dimensional version does not.

	Moreover, we further explored the Carroll-Schr\"odinger field theory by applying the method of canonical quantization and computing the transition amplitude \eqref{tam}. The latter corresponded with the results in \cite{Afshar:2024llh}.

	Furthermore, we attempted to extend our results to higher dimensions. By considering our initial framework in higher dimensions and applying the Carroll limit, we derived an equation in higher dimensions, referred to as the generalized Carroll-Schr\"odinger equation \eqref{inany}. In two dimensions, this equation appropriately reduces to the Carroll-Schr\"odinger equation \eqref{cse}.

	It becomes an intriguing question to explore the possibilities of extending the Carroll-Schr\"odinger algebra beyond two dimensions. In arbitrary dimensions, we found that the Jacobi identity cannot be satisfied, except in two dimensions. This suggests that extending the algebra \eqref{csch} to higher dimensions may require activating some of the previously vanished commutators. Such an extension to higher dimensions may serve as the symmetry algebra for the generalized Carroll-Schr\"odinger equation \eqref{inany}, a task we plan to explore in the future.

	The action \eqref{inanya} demonstrates a global phase symmetry under the transformation $\psi\to e^{i\lambda}\,\psi$. Therefore, exploring a local phase symmetry could present an interesting research problem.

	Moreover, as we know, the Schr\"odinger equation can also be derived from the Dirac equation (see e.g. \cite{Leblanc:1992wu}). Consequently, it becomes interesting to explore the possibility of deriving the Carroll-Schr\"odinger equation \eqref{cse} from a tachyonic Dirac equation.

	In both Galilean and Carrollian field theories, it is known that there are two distinct sectors: the electric sector and the magnetic sector. A comprehensive review of electric and magnetic scalar fields, as well as electric and magnetic spinor fields, can be found in \cite{Koutrolikos:2023evq}. It is noteworthy that both the electric and magnetic versions share the same underlying symmetry algebra, whether it be Galilean or Carrollian.

	With this in mind, if we refer to the Schr\"odinger equation as the electric version, it raises the question of what the magnetic version of the Schr\"odinger equation would be. Similarly, if we consider the obtained Carroll-Schr\"odinger equation \eqref{cse} as the magnetic version, we may inquire about its electric counterpart. One possible approach to deriving such equations is through the utilization of the seed Lagrangian method, presented in \cite{Bergshoeff:2022qkx} (for further details, refer also to \cite{Koutrolikos:2023evq}).

	The Carroll-Schr\"odinger equation \eqref{cse} and the generalized equation \eqref{inany} can provide a promising avenue for the development of Carrollian quantum mechanics, with potential applications in condensed matter systems. This could open up new possibilities for addressing various problems within the Carrollian framework, such as: the infinite potential well, the Aharanov B\"ohm Effect, the Hydrogen atom, the harmonic oscillator and so on. As we were finalizing this work, we came across this reference \cite{Vysin:1977ue} that derived the two-dimensional equation \eqref{cse} by a different method. It also solved the problem of the infinite potential well.

	By solving the equation for different potential energy scenarios, we can compare the results with established outcomes and examine any differences. Notably, it seems the presence of the speed of light $c$ in the Carroll-Schr\"odinger equation \eqref{cse} has posed challenges in finding the effects of Carrollian quantum mechanics so far. However, it is anticipated that a careful analysis may reveal potential shifts in energy levels associated with this novel framework.

	
	

	\noindent{\bf Acknowledgments:} We are grateful to Hamid Afshar and Ahmad Ghodsi for useful discussions and comments, and to Shahin Sheikh-Jabbari for his support. We also thank the referees for their valuable suggestions, which have improved the quality of this paper. This work is based upon research funded by Iran National Science Foundation (INSF) under project No. 4028530. The author is also partially supported by IPM funds.
	

	\appendix

	\onecolumngrid
	
	
	\begin{center}
		\noindent\rule{510pt}{.5pt}	
	\end{center}
	
	\section{Schr\"odinger equation} \label{ap}
	Let us consider the Klein–Gordon equation for a complex scalar field $\upphi$ of mass $m$ in a $(d+1)$-dimensional Minkowski spacetime, with the mostly plus signature for the metric, including the speed of light $c$ and the reduced Planck constant $\hbar$
	\begin{align}
		\le(-\,\frac{1}{c^2}\,\p_t^2\,+\,\p^i\p_i\,-\,\m^2 \ri){\upphi}=0\,,\label{KG1}
	\end{align} 
	where $\m:={mc}/{\hbar}$. Using a field redefinition
	\begin{align}
		\upphi=\frac{1}{\sqrt{\m}}~e^{-i\m ct}~\uppsi\,, \label{fieldred1}
	\end{align}
	the Klein-Gordon equation \eqref{KG1} reduces to
	\begin{align}
		\le(-\,\frac{1}{c^2}\,\p_t^2\,+\,\frac{2im}{\hbar}\,\p_t\,+\,\p^i\p_i\ri)\uppsi=0\,.\label{reduced1}
	\end{align}
	By applying the Galilei or nonrelativistic limit ($c\to\infty$) to the latter, and multiplying by $\hbar^2/2m$, we will arrive at the Schr\"odinger equation
	\begin{align}
		\le(i\hbar\,\p_t\,+\,\frac{\hbar^2}{2m}\,\p^i\p_i\ri)\uppsi=0\,.\label{sch1}
	\end{align} 
	This equation can be obtain from the Schr\"odinger action
	\begin{align}
		S=\int dt\,d^dx~\uppsi^\star\le(i\hbar\,\p_t\,+\,\frac{\hbar^2}{2m}\,\p^i\p_i\ri)\uppsi\,.\label{sch1a}
	\end{align} 
	We note that the length dimension of a scalar field is $[\upphi]_{_L}=(1-d)/2$. Therefore, following \eqref{fieldred1}, the length dimension for the Schr\"odinger field $\uppsi$ reads $[\uppsi]_{_L}=-\,d/2$, and as a result the action \eqref{sch1a} possesses the expected dimension of $\hbar$.



\newpage
\twocolumngrid

\bibliographystyle{hephys}

\bibliography{references}

\begin{thebibliography}{10}
\newcommand{\enquote}[1]{``#1''}

\bibitem{Levy1965}
J.-M. Lévy-Leblond, \enquote{Une nouvelle limite non-relativiste du groupe de
  Poincaré}, \href{http://eudml.org/doc/75509}{\emph{Annales de l'I.H.P.
  Physique théorique} \textbf{3[1]} (1965) 1}.

\bibitem{Duval:2014uoa}
C.~Duval, G.~W. Gibbons, P.~A. Horvathy and P.~M. Zhang, \enquote{{Carroll
  versus Newton and Galilei: two dual non-Einsteinian concepts of time}},
  \href{http://dx.doi.org/10.1088/0264-9381/31/8/085016}{\emph{Class. Quant.
  Grav.} \textbf{31} (2014) 085016}, \href{http://arxiv.org/abs/1402.0657}{{\tt
  arXiv:1402.0657 [gr-qc]}}.

\bibitem{Duval:2014uva}
C.~Duval, G.~W. Gibbons and P.~A. Horvathy, \enquote{{Conformal Carroll groups
  and BMS symmetry}},
  \href{http://dx.doi.org/10.1088/0264-9381/31/9/092001}{\emph{Class. Quant.
  Grav.} \textbf{31} (2014) 092001}, \href{http://arxiv.org/abs/1402.5894}{{\tt
  arXiv:1402.5894 [gr-qc]}}.

\bibitem{Duval:2014lpa}
C.~Duval, G.~W. Gibbons and P.~A. Horvathy, \enquote{{Conformal Carroll
  groups}}, \href{http://dx.doi.org/10.1088/1751-8113/47/33/335204}{\emph{J.
  Phys. A} \textbf{47[33]} (2014) 335204},
  \href{http://arxiv.org/abs/1403.4213}{{\tt arXiv:1403.4213 [hep-th]}}.

\bibitem{Barnich:2009se}
G.~Barnich and C.~Troessaert, \enquote{{Symmetries of asymptotically flat 4
  dimensional spacetimes at null infinity revisited}},
  \href{http://dx.doi.org/10.1103/PhysRevLett.105.111103}{\emph{Phys. Rev.
  Lett.} \textbf{105} (2010) 111103},
  \href{http://arxiv.org/abs/0909.2617}{{\tt arXiv:0909.2617 [gr-qc]}}.

\bibitem{Bagchi:2010zz}
A.~Bagchi, \enquote{{Correspondence between Asymptotically Flat Spacetimes and
  Nonrelativistic Conformal Field Theories}},
  \href{http://dx.doi.org/10.1103/PhysRevLett.105.171601}{\emph{Phys. Rev.
  Lett.} \textbf{105} (2010) 171601},
  \href{http://arxiv.org/abs/1006.3354}{{\tt arXiv:1006.3354 [hep-th]}}.

\bibitem{Ciambelli:2018wre}
L.~Ciambelli, C.~Marteau, A.~C. Petkou, P.~M. Petropoulos and K.~Siampos,
  \enquote{{Flat holography and Carrollian fluids}},
  \href{http://dx.doi.org/10.1007/JHEP07(2018)165}{\emph{JHEP} \textbf{07}
  (2018) 165}, \href{http://arxiv.org/abs/1802.06809}{{\tt arXiv:1802.06809
  [hep-th]}}.

\bibitem{Figueroa-OFarrill:2021sxz}
J.~Figueroa-O'Farrill, E.~Have, S.~Prohazka and J.~Salzer, \enquote{{Carrollian
  and celestial spaces at infinity}},
  \href{http://dx.doi.org/10.1007/JHEP09(2022)007}{\emph{JHEP} \textbf{09}
  (2022) 007}, \href{http://arxiv.org/abs/2112.03319}{{\tt arXiv:2112.03319
  [hep-th]}}.

\bibitem{Herfray:2021qmp}
Y.~Herfray, \enquote{{Carrollian manifolds and null infinity: a view from
  Cartan geometry}},
  \href{http://dx.doi.org/10.1088/1361-6382/ac635f}{\emph{Class. Quant. Grav.}
  \textbf{39[21]} (2022) 215005}, \href{http://arxiv.org/abs/2112.09048}{{\tt
  arXiv:2112.09048 [gr-qc]}}.

\bibitem{Bagchi:2022emh}
A.~Bagchi, S.~Banerjee, R.~Basu and S.~Dutta, \enquote{{Scattering Amplitudes:
  Celestial and Carrollian}},
  \href{http://dx.doi.org/10.1103/PhysRevLett.128.241601}{\emph{Phys. Rev.
  Lett.} \textbf{128[24]} (2022) 241601},
  \href{http://arxiv.org/abs/2202.08438}{{\tt arXiv:2202.08438 [hep-th]}}.

\bibitem{Donnay:2022aba}
L.~Donnay, A.~Fiorucci, Y.~Herfray and R.~Ruzziconi, \enquote{{Carrollian
  Perspective on Celestial Holography}},
  \href{http://dx.doi.org/10.1103/PhysRevLett.129.071602}{\emph{Phys. Rev.
  Lett.} \textbf{129[7]} (2022) 071602},
  \href{http://arxiv.org/abs/2202.04702}{{\tt arXiv:2202.04702 [hep-th]}}.

\bibitem{Campoleoni:2022wmf}
A.~Campoleoni, L.~Ciambelli, A.~Delfante, C.~Marteau, P.~M. Petropoulos and
  R.~Ruzziconi, \enquote{{Holographic Lorentz and Carroll frames}},
  \href{http://dx.doi.org/10.1007/JHEP12(2022)007}{\emph{JHEP} \textbf{12}
  (2022) 007}, \href{http://arxiv.org/abs/2208.07575}{{\tt arXiv:2208.07575
  [hep-th]}}.

\bibitem{Donnay:2022wvx}
L.~Donnay, A.~Fiorucci, Y.~Herfray and R.~Ruzziconi, \enquote{{Bridging
  Carrollian and celestial holography}},
  \href{http://dx.doi.org/10.1103/PhysRevD.107.126027}{\emph{Phys. Rev. D}
  \textbf{107[12]} (2023) 126027}, \href{http://arxiv.org/abs/2212.12553}{{\tt
  arXiv:2212.12553 [hep-th]}}.

\bibitem{Bagchi:2023fbj}
A.~Bagchi, P.~Dhivakar and S.~Dutta, \enquote{{AdS Witten diagrams to
  Carrollian correlators}},
  \href{http://dx.doi.org/10.1007/JHEP04(2023)135}{\emph{JHEP} \textbf{04}
  (2023) 135}, \href{http://arxiv.org/abs/2303.07388}{{\tt arXiv:2303.07388
  [hep-th]}}.

\bibitem{Saha:2023hsl}
A.~Saha, \enquote{{Carrollian approach to 1 + 3D flat holography}},
  \href{http://dx.doi.org/10.1007/JHEP06(2023)051}{\emph{JHEP} \textbf{06}
  (2023) 051}, \href{http://arxiv.org/abs/2304.02696}{{\tt arXiv:2304.02696
  [hep-th]}}.

\bibitem{Salzer:2023jqv}
J.~Salzer, \enquote{{An embedding space approach to Carrollian CFT correlators
  for flat space holography}},
  \href{http://dx.doi.org/10.1007/JHEP10(2023)084}{\emph{JHEP} \textbf{10}
  (2023) 084}, \href{http://arxiv.org/abs/2304.08292}{{\tt arXiv:2304.08292
  [hep-th]}}.

\bibitem{Campoleoni:2023fug}
A.~Campoleoni, A.~Delfante, S.~Pekar, P.~M. Petropoulos, D.~Rivera-Betancour
  and M.~Vilatte, \enquote{{Flat from anti de Sitter}},
  \href{http://dx.doi.org/10.1007/JHEP12(2023)078}{\emph{JHEP} \textbf{12}
  (2023) 078}, \href{http://arxiv.org/abs/2309.15182}{{\tt arXiv:2309.15182
  [hep-th]}}.

\bibitem{Nguyen:2023vfz}
K.~Nguyen and P.~West, \enquote{{Carrollian Conformal Fields and Flat
  Holography}},
  \href{http://dx.doi.org/10.3390/universe9090385}{\emph{Universe}
  \textbf{9[9]} (2023) 385}, \href{http://arxiv.org/abs/2305.02884}{{\tt
  arXiv:2305.02884 [hep-th]}}.

\bibitem{Henneaux:1979vn}
M.~Henneaux, \enquote{{Geometry of Zero Signature Space-times}}, \emph{Bull.
  Soc. Math. Belg.} \textbf{31} (1979) 47.

\bibitem{Dautcourt:1997hb}
G.~Dautcourt, \enquote{{On the ultrarelativistic limit of general relativity}},
  \emph{Acta Phys. Polon. B} \textbf{29} (1998) 1047,
  \href{http://arxiv.org/abs/gr-qc/9801093}{{\tt arXiv:gr-qc/9801093}}.

\bibitem{Bergshoeff:2014jla}
E.~Bergshoeff, J.~Gomis and G.~Longhi, \enquote{{Dynamics of Carroll
  Particles}},
  \href{http://dx.doi.org/10.1088/0264-9381/31/20/205009}{\emph{Class. Quant.
  Grav.} \textbf{31[20]} (2014) 205009},
  \href{http://arxiv.org/abs/1405.2264}{{\tt arXiv:1405.2264 [hep-th]}}.

\bibitem{Nzotungicimpaye:2014wya}
J.~Nzotungicimpaye, \enquote{{Kinematical versus Dynamical Contractions of the
  de Sitter Lie algebras}},
  \href{http://dx.doi.org/10.1088/2399-6528/ab4683}{\emph{J. Phys. Comm.}
  \textbf{3[10]} (2019) 105003}, \href{http://arxiv.org/abs/1406.0972}{{\tt
  arXiv:1406.0972 [math-ph]}}.

\bibitem{Bergshoeff:2015wma}
E.~Bergshoeff, J.~Gomis and L.~Parra, \enquote{{The Symmetries of the Carroll
  Superparticle}},
  \href{http://dx.doi.org/10.1088/1751-8113/49/18/185402}{\emph{J. Phys. A}
  \textbf{49[18]} (2016) 185402}, \href{http://arxiv.org/abs/1503.06083}{{\tt
  arXiv:1503.06083 [hep-th]}}.

\bibitem{Bekaert:2015xua}
X.~Bekaert and K.~Morand, \enquote{{Connections and dynamical trajectories in
  generalised Newton-Cartan gravity II. An ambient perspective}},
  \href{http://dx.doi.org/10.1063/1.5030328}{\emph{J. Math. Phys.}
  \textbf{59[7]} (2018) 072503}, \href{http://arxiv.org/abs/1505.03739}{{\tt
  arXiv:1505.03739 [hep-th]}}.

\bibitem{Hartong:2015xda}
J.~Hartong, \enquote{{Gauging the Carroll Algebra and Ultra-Relativistic
  Gravity}}, \href{http://dx.doi.org/10.1007/JHEP08(2015)069}{\emph{JHEP}
  \textbf{08} (2015) 069}, \href{http://arxiv.org/abs/1505.05011}{{\tt
  arXiv:1505.05011 [hep-th]}}.

\bibitem{Bergshoeff:2017btm}
E.~Bergshoeff, J.~Gomis, B.~Rollier, J.~Rosseel and T.~ter Veldhuis,
  \enquote{{Carroll versus Galilei Gravity}},
  \href{http://dx.doi.org/10.1007/JHEP03(2017)165}{\emph{JHEP} \textbf{03}
  (2017) 165}, \href{http://arxiv.org/abs/1701.06156}{{\tt arXiv:1701.06156
  [hep-th]}}.

\bibitem{Duval:2017els}
C.~Duval, G.~W. Gibbons, P.~A. Horvathy and P.~M. Zhang, \enquote{{Carroll
  symmetry of plane gravitational waves}},
  \href{http://dx.doi.org/10.1088/1361-6382/aa7f62}{\emph{Class. Quant. Grav.}
  \textbf{34[17]} (2017) 175003}, \href{http://arxiv.org/abs/1702.08284}{{\tt
  arXiv:1702.08284 [gr-qc]}}.

\bibitem{deBoer:2017ing}
J.~de~Boer, J.~Hartong, N.~A. Obers, W.~Sybesma and S.~Vandoren,
  \enquote{{Perfect Fluids}},
  \href{http://dx.doi.org/10.21468/SciPostPhys.5.1.003}{\emph{SciPost Phys.}
  \textbf{5[1]} (2018) 003}, \href{http://arxiv.org/abs/1710.04708}{{\tt
  arXiv:1710.04708 [hep-th]}}.

\bibitem{Ciambelli:2018xat}
L.~Ciambelli, C.~Marteau, A.~C. Petkou, P.~M. Petropoulos and K.~Siampos,
  \enquote{{Covariant Galilean versus Carrollian hydrodynamics from
  relativistic fluids}},
  \href{http://dx.doi.org/10.1088/1361-6382/aacf1a}{\emph{Class. Quant. Grav.}
  \textbf{35[16]} (2018) 165001}, \href{http://arxiv.org/abs/1802.05286}{{\tt
  arXiv:1802.05286 [hep-th]}}.

\bibitem{Ciambelli:2018ojf}
L.~Ciambelli and C.~Marteau, \enquote{{Carrollian conservation laws and
  Ricci-flat gravity}},
  \href{http://dx.doi.org/10.1088/1361-6382/ab0d37}{\emph{Class. Quant. Grav.}
  \textbf{36[8]} (2019) 085004}, \href{http://arxiv.org/abs/1810.11037}{{\tt
  arXiv:1810.11037 [hep-th]}}.

\bibitem{Morand:2018tke}
K.~Morand, \enquote{{Embedding Galilean and Carrollian geometries I.
  Gravitational waves}}, \href{http://dx.doi.org/10.1063/1.5130907}{\emph{J.
  Math. Phys.} \textbf{61[8]} (2020) 082502},
  \href{http://arxiv.org/abs/1811.12681}{{\tt arXiv:1811.12681 [hep-th]}}.

\bibitem{Penna:2018gfx}
R.~F. Penna, \enquote{{Near-horizon Carroll symmetry and black hole Love
  numbers}}, \href{http://arxiv.org/abs/1812.05643}{{\tt arXiv:1812.05643
  [hep-th]}}.

\bibitem{Donnay:2019jiz}
L.~Donnay and C.~Marteau, \enquote{{Carrollian Physics at the Black Hole
  Horizon}}, \href{http://dx.doi.org/10.1088/1361-6382/ab2fd5}{\emph{Class.
  Quant. Grav.} \textbf{36[16]} (2019) 165002},
  \href{http://arxiv.org/abs/1903.09654}{{\tt arXiv:1903.09654 [hep-th]}}.

\bibitem{Bergshoeff:2019ctr}
E.~Bergshoeff, J.~M. Izquierdo, T.~Ort\'\i{}n and L.~Romano, \enquote{{Lie
  Algebra Expansions and Actions for Non-Relativistic Gravity}},
  \href{http://dx.doi.org/10.1007/JHEP08(2019)048}{\emph{JHEP} \textbf{08}
  (2019) 048}, \href{http://arxiv.org/abs/1904.08304}{{\tt arXiv:1904.08304
  [hep-th]}}.

\bibitem{Ravera:2019ize}
L.~Ravera, \enquote{{AdS Carroll Chern-Simons supergravity in 2 + 1 dimensions
  and its flat limit}},
  \href{http://dx.doi.org/10.1016/j.physletb.2019.06.026}{\emph{Phys. Lett. B}
  \textbf{795} (2019) 331}, \href{http://arxiv.org/abs/1905.00766}{{\tt
  arXiv:1905.00766 [hep-th]}}.

\bibitem{Gomis:2019nih}
J.~Gomis, A.~Kleinschmidt, J.~Palmkvist and P.~Salgado-Rebolledo,
  \enquote{{Newton-Hooke/Carrollian expansions of (A)dS and Chern-Simons
  gravity}}, \href{http://dx.doi.org/10.1007/JHEP02(2020)009}{\emph{JHEP}
  \textbf{02} (2020) 009}, \href{http://arxiv.org/abs/1912.07564}{{\tt
  arXiv:1912.07564 [hep-th]}}.

\bibitem{Ciambelli:2019lap}
L.~Ciambelli, R.~G. Leigh, C.~Marteau and P.~M. Petropoulos, \enquote{{Carroll
  Structures, Null Geometry and Conformal Isometries}},
  \href{http://dx.doi.org/10.1103/PhysRevD.100.046010}{\emph{Phys. Rev. D}
  \textbf{100[4]} (2019) 046010}, \href{http://arxiv.org/abs/1905.02221}{{\tt
  arXiv:1905.02221 [hep-th]}}.

\bibitem{Ballesteros:2019mxi}
A.~Ballesteros, G.~Gubitosi and F.~J. Herranz, \enquote{{Lorentzian Snyder
  spacetimes and their Galilei and Carroll limits from projective geometry}},
  \href{http://dx.doi.org/10.1088/1361-6382/aba668}{\emph{Class. Quant. Grav.}
  \textbf{37[19]} (2020) 195021}, \href{http://arxiv.org/abs/1912.12878}{{\tt
  arXiv:1912.12878 [hep-th]}}.

\bibitem{Bergshoeff:2020xhv}
E.~Bergshoeff, J.~M. Izquierdo and L.~Romano, \enquote{{Carroll versus Galilei
  from a Brane Perspective}},
  \href{http://dx.doi.org/10.1007/JHEP10(2020)066}{\emph{JHEP} \textbf{10}
  (2020) 066}, \href{http://arxiv.org/abs/2003.03062}{{\tt arXiv:2003.03062
  [hep-th]}}.

\bibitem{Niedermaier:2020jdy}
M.~Niedermaier, \enquote{{Nonstandard Action of Diffeomorphisms and
  Gravity\textquoteright{}s Anti-Newtonian Limit}},
  \href{http://dx.doi.org/10.3390/sym12050752}{\emph{Symmetry} \textbf{12[5]}
  (2020) 752}.

\bibitem{Gomis:2020wxp}
J.~Gomis, D.~Hidalgo and P.~Salgado-Rebolledo, \enquote{{Non-relativistic and
  Carrollian limits of Jackiw-Teitelboim gravity}},
  \href{http://dx.doi.org/10.1007/JHEP05(2021)162}{\emph{JHEP} \textbf{05}
  (2021) 162}, \href{http://arxiv.org/abs/2011.15053}{{\tt arXiv:2011.15053
  [hep-th]}}.

\bibitem{Grumiller:2020elf}
D.~Grumiller, J.~Hartong, S.~Prohazka and J.~Salzer, \enquote{{Limits of JT
  gravity}}, \href{http://dx.doi.org/10.1007/JHEP02(2021)134}{\emph{JHEP}
  \textbf{02} (2021) 134}, \href{http://arxiv.org/abs/2011.13870}{{\tt
  arXiv:2011.13870 [hep-th]}}.

\bibitem{Hansen:2021fxi}
D.~Hansen, N.~A. Obers, G.~Oling and B.~T. S\o{}gaard, \enquote{{Carroll
  Expansion of General Relativity}},
  \href{http://dx.doi.org/10.21468/SciPostPhys.13.3.055}{\emph{SciPost Phys.}
  \textbf{13[3]} (2022) 055}, \href{http://arxiv.org/abs/2112.12684}{{\tt
  arXiv:2112.12684 [hep-th]}}.

\bibitem{deBoer:2021jej}
J.~de~Boer, J.~Hartong, N.~A. Obers, W.~Sybesma and S.~Vandoren,
  \enquote{{Carroll Symmetry, Dark Energy and Inflation}},
  \href{http://dx.doi.org/10.3389/fphy.2022.810405}{\emph{Front. in Phys.}
  \textbf{10} (2022) 810405}, \href{http://arxiv.org/abs/2110.02319}{{\tt
  arXiv:2110.02319 [hep-th]}}.

\bibitem{Henneaux:2021yzg}
M.~Henneaux and P.~Salgado-Rebolledo, \enquote{{Carroll contractions of
  Lorentz-invariant theories}},
  \href{http://dx.doi.org/10.1007/JHEP11(2021)180}{\emph{JHEP} \textbf{11}
  (2021) 180}, \href{http://arxiv.org/abs/2109.06708}{{\tt arXiv:2109.06708
  [hep-th]}}.

\bibitem{Concha:2021jnn}
P.~Concha, D.~Pe\~nafiel, L.~Ravera and E.~Rodr\'\i{}guez,
  \enquote{{Three-dimensional Maxwellian Carroll gravity theory and the
  cosmological constant}},
  \href{http://dx.doi.org/10.1016/j.physletb.2021.136735}{\emph{Phys. Lett. B}
  \textbf{823} (2021) 136735}, \href{http://arxiv.org/abs/2107.05716}{{\tt
  arXiv:2107.05716 [hep-th]}}.

\bibitem{Guerrieri:2021cdz}
A.~Guerrieri and R.~F. Sobreiro, \enquote{{Carroll limit of four-dimensional
  gravity theories in the first order formalism}},
  \href{http://dx.doi.org/10.1088/1361-6382/ac345f}{\emph{Class. Quant. Grav.}
  \textbf{38[24]} (2021) 245003}, \href{http://arxiv.org/abs/2107.10129}{{\tt
  arXiv:2107.10129 [gr-qc]}}.

\bibitem{Perez:2021abf}
A.~P\'erez, \enquote{{Asymptotic symmetries in Carrollian theories of
  gravity}}, \href{http://dx.doi.org/10.1007/JHEP12(2021)173}{\emph{JHEP}
  \textbf{12} (2021) 173}, \href{http://arxiv.org/abs/2110.15834}{{\tt
  arXiv:2110.15834 [hep-th]}}.

\bibitem{Figueroa-OFarrill:2022mcy}
J.~Figueroa-O'Farrill, E.~Have, S.~Prohazka and J.~Salzer, \enquote{{The
  gauging procedure and carrollian gravity}},
  \href{http://dx.doi.org/10.1007/JHEP09(2022)243}{\emph{JHEP} \textbf{09}
  (2022) 243}, \href{http://arxiv.org/abs/2206.14178}{{\tt arXiv:2206.14178
  [hep-th]}}.

\bibitem{Campoleoni:2022ebj}
A.~Campoleoni, M.~Henneaux, S.~Pekar, A.~P\'erez and P.~Salgado-Rebolledo,
  \enquote{{Magnetic Carrollian gravity from the Carroll algebra}},
  \href{http://dx.doi.org/10.1007/JHEP09(2022)127}{\emph{JHEP} \textbf{09}
  (2022) 127}, \href{http://arxiv.org/abs/2207.14167}{{\tt arXiv:2207.14167
  [hep-th]}}.

\bibitem{Baiguera:2022lsw}
S.~Baiguera, G.~Oling, W.~Sybesma and B.~T. S\o{}gaard, \enquote{{Conformal
  Carroll scalars with boosts}},
  \href{http://dx.doi.org/10.21468/SciPostPhys.14.4.086}{\emph{SciPost Phys.}
  \textbf{14[4]} (2023) 086}, \href{http://arxiv.org/abs/2207.03468}{{\tt
  arXiv:2207.03468 [hep-th]}}.

\bibitem{Perez:2022jpr}
A.~P\'erez, \enquote{{Asymptotic symmetries in Carrollian theories of gravity
  with a negative cosmological constant}},
  \href{http://dx.doi.org/10.1007/JHEP09(2022)044}{\emph{JHEP} \textbf{09}
  (2022) 044}, \href{http://arxiv.org/abs/2202.08768}{{\tt arXiv:2202.08768
  [hep-th]}}.

\bibitem{Fuentealba:2022gdx}
O.~Fuentealba, M.~Henneaux, P.~Salgado-Rebolledo and J.~Salzer,
  \enquote{{Asymptotic structure of Carrollian limits of Einstein-Yang-Mills
  theory in four spacetime dimensions}},
  \href{http://dx.doi.org/10.1103/PhysRevD.106.104047}{\emph{Phys. Rev. D}
  \textbf{106[10]} (2022) 104047}, \href{http://arxiv.org/abs/2207.11359}{{\tt
  arXiv:2207.11359 [hep-th]}}.

\bibitem{Marsot:2022imf}
L.~Marsot, P.~M. Zhang, M.~Chernodub and P.~A. Horvathy, \enquote{{Hall effects
  in Carroll dynamics}},
  \href{http://dx.doi.org/10.1016/j.physrep.2023.07.007}{\emph{Phys. Rept.}
  \textbf{1028} (2023) 1}, \href{http://arxiv.org/abs/2212.02360}{{\tt
  arXiv:2212.02360 [hep-th]}}.

\bibitem{Bergshoeff:2022qkx}
E.~A. Bergshoeff, J.~Gomis and A.~Kleinschmidt, \enquote{{Non-Lorentzian
  theories with and without constraints}},
  \href{http://dx.doi.org/10.1007/JHEP01(2023)167}{\emph{JHEP} \textbf{01}
  (2023) 167}, \href{http://arxiv.org/abs/2210.14848}{{\tt arXiv:2210.14848
  [hep-th]}}.

\bibitem{Banerjee:2022ocj}
A.~Banerjee, S.~Dutta and S.~Mondal, \enquote{{Carroll fermions in two
  dimensions}},
  \href{http://dx.doi.org/10.1103/PhysRevD.107.125020}{\emph{Phys. Rev. D}
  \textbf{107[12]} (2023) 125020}, \href{http://arxiv.org/abs/2211.11639}{{\tt
  arXiv:2211.11639 [hep-th]}}.

\bibitem{Bagchi:2022eui}
A.~Bagchi, A.~Banerjee, R.~Basu, M.~Islam and S.~Mondal, \enquote{{Magic
  fermions: Carroll and flat bands}},
  \href{http://dx.doi.org/10.1007/JHEP03(2023)227}{\emph{JHEP} \textbf{03}
  (2023) 227}, \href{http://arxiv.org/abs/2211.11640}{{\tt arXiv:2211.11640
  [hep-th]}}.

\bibitem{Bagchi:2022owq}
A.~Bagchi, D.~Grumiller and P.~Nandi, \enquote{{Carrollian superconformal
  theories and super BMS}},
  \href{http://dx.doi.org/10.1007/JHEP05(2022)044}{\emph{JHEP} \textbf{05}
  (2022) 044}, \href{http://arxiv.org/abs/2202.01172}{{\tt arXiv:2202.01172
  [hep-th]}}.

\bibitem{Bekaert:2022oeh}
X.~Bekaert, A.~Campoleoni and S.~Pekar, \enquote{{Carrollian conformal scalar
  as flat-space singleton}},
  \href{http://dx.doi.org/10.1016/j.physletb.2023.137734}{\emph{Phys. Lett. B}
  \textbf{838} (2023) 137734}, \href{http://arxiv.org/abs/2211.16498}{{\tt
  arXiv:2211.16498 [hep-th]}}.

\bibitem{Rivera-Betancour:2022lkc}
D.~Rivera-Betancour and M.~Vilatte, \enquote{{Revisiting the Carrollian scalar
  field}}, \href{http://dx.doi.org/10.1103/PhysRevD.106.085004}{\emph{Phys.
  Rev. D} \textbf{106[8]} (2022) 085004},
  \href{http://arxiv.org/abs/2207.01647}{{\tt arXiv:2207.01647 [hep-th]}}.

\bibitem{deBoer:2023fnj}
J.~de~Boer, J.~Hartong, N.~A. Obers, W.~Sybesma and S.~Vandoren,
  \enquote{{Carroll stories}},
  \href{http://dx.doi.org/10.1007/JHEP09(2023)148}{\emph{JHEP} \textbf{09}
  (2023) 148}, \href{http://arxiv.org/abs/2307.06827}{{\tt arXiv:2307.06827
  [hep-th]}}.

\bibitem{Ecker:2023uwm}
F.~Ecker, D.~Grumiller, J.~Hartong, A.~P\'erez, S.~Prohazka and R.~Troncoso,
  \enquote{{Carroll black holes}},
  \href{http://dx.doi.org/10.21468/SciPostPhys.15.6.245}{\emph{SciPost Phys.}
  \textbf{15[6]} (2023) 245}, \href{http://arxiv.org/abs/2308.10947}{{\tt
  arXiv:2308.10947 [hep-th]}}.

\bibitem{Koutrolikos:2023evq}
K.~Koutrolikos and M.~Najafizadeh, \enquote{{Super-Carrollian and
  Super-Galilean Field Theories}},
  \href{http://dx.doi.org/10.1103/PhysRevD.108.125014}{\emph{Phys. Rev. D}
  \textbf{108[12]} (2023) 125014}, \href{http://arxiv.org/abs/2309.16786}{{\tt
  arXiv:2309.16786 [hep-th]}}.

\bibitem{Kasikci:2023zdn}
O.~Kasikci, M.~Ozkan, Y.~Pang and U.~Zorba, \enquote{{Carrollian Supersymmetry
  and SYK-like models}}, \href{http://arxiv.org/abs/2311.00039}{{\tt
  arXiv:2311.00039 [hep-th]}}.

\bibitem{Ciambelli:2023tzb}
L.~Ciambelli and D.~Grumiller, \enquote{{Carroll geodesics}},
  \href{http://arxiv.org/abs/2311.04112}{{\tt arXiv:2311.04112 [hep-th]}}.

\bibitem{Ciambelli:2023xqk}
L.~Ciambelli, \enquote{{Dynamics of Carrollian Scalar Fields}},
  \href{http://arxiv.org/abs/2311.04113}{{\tt arXiv:2311.04113 [hep-th]}}.

\bibitem{Bergshoeff:2023vfd}
E.~A. Bergshoeff, A.~Campoleoni, A.~Fontanella, L.~Mele and J.~Rosseel,
  \enquote{{Carroll Fermions}}, \href{http://arxiv.org/abs/2312.00745}{{\tt
  arXiv:2312.00745 [hep-th]}}.

\bibitem{Leblanc:1992wu}
M.~Leblanc, G.~Lozano and H.~Min, \enquote{{Extended superconformal Galilean
  symmetry in Chern-Simons matter systems}},
  \href{http://dx.doi.org/10.1016/0003-4916(92)90350-U}{\emph{Annals Phys.}
  \textbf{219} (1992) 328}, \href{http://arxiv.org/abs/hep-th/9206039}{{\tt
  arXiv:hep-th/9206039}}.

\bibitem{Jackiw:1990mb}
R.~Jackiw and S.-Y. Pi, \enquote{{Classical and quantal nonrelativistic
  Chern-Simons theory}},
  \href{http://dx.doi.org/10.1103/PhysRevD.42.3500}{\emph{Phys. Rev. D}
  \textbf{42} (1990) 3500}, [Erratum: Phys.Rev.D 48, 3929 (1993)].

\bibitem{Henkel:1993sg}
M.~Henkel, \enquote{{Schrodinger invariance in strongly anisotropic critical
  systems}}, \href{http://dx.doi.org/10.1007/BF02186756}{\emph{J. Statist.
  Phys.} \textbf{75} (1994) 1023},
  \href{http://arxiv.org/abs/hep-th/9310081}{{\tt arXiv:hep-th/9310081}}.

\bibitem{Nishida:2007pj}
Y.~Nishida and D.~T. Son, \enquote{{Nonrelativistic conformal field theories}},
  \href{http://dx.doi.org/10.1103/PhysRevD.76.086004}{\emph{Phys. Rev. D}
  \textbf{76} (2007) 086004}, \href{http://arxiv.org/abs/0706.3746}{{\tt
  arXiv:0706.3746 [hep-th]}}.

\bibitem{Vysin:1977ue}
V.~Vysin, \enquote{{Nonrelativistic Reduction and Interpretation of the
  Klein-Gordon Equation of Tachyons}},
  \href{http://dx.doi.org/10.1007/BF02776778}{\emph{Nuovo Cim. A} \textbf{40}
  (1977) 113}.

\bibitem{Najafizadeh:2024}
M.~Najafizadeh, \enquote{work in progress}, .

\bibitem{Afshar:2024llh}
H.~Afshar, X.~Bekaert and M.~Najafizadeh, \enquote{{Classification of Conformal
  Carroll Algebras}}, \href{http://arxiv.org/abs/2409.19953}{{\tt
  arXiv:2409.19953 [hep-th]}}.

\end{thebibliography}

\end{document}